\documentstyle[12pt,psfig]{article}

\def\baselinestretch{1.2}
\def\href#1#2{#2}

\newcommand{\norm}[1]{\raise.3ex\hbox{:} #1 \raise.3ex\hbox{:}\,}

\newcommand{\beq}{\begin{equation}}
\newcommand{\eeq}{\end{equation}}

\def\appendix{{\newpage\section*{Appendix}}\let\appendix\section%
        {\setcounter{section}{0}
        \gdef\thesection{\Alph{section}}}\section}

\begin{document}

\begin{titlepage}

\begin{flushright}
NSF-ITP-99-040\\
OHSTPY-HEP-T-99-012\\
hep-th/9906087
\end{flushright}
\vfil\vfil

\begin{center}

{\Large {\bf Can DLCQ test  the Maldacena Conjecture?}}

\vfil

Francesco Antonuccio$^a$, Akikazu Hashimoto$^b$, Oleg Lunin$^a$, and
Stephen Pinsky$^a$\\

\vfil

$^a$Department of Physics \\ Ohio State University, Columbus, Ohio 43210\\

\vfil

$^b$Institute for Theoretical Physics\\ University of California,
Santa Barbara, CA 93106\\

\vfil
\end{center}

\begin{abstract}
\noindent We consider the Maldacena conjecture applied to the near
horizon geometry of a D1-brane in the supergravity approximation and
consider the possibility of testing the conjecture against the
boundary field theory calculation using DLCQ.  We propose the two
point function of the stress energy tensor as a convenient quantity
that may be computed on both sides of the correspondence. On the
supergravity side, we may invoke the methods of Gubser, Klebanov,
Polyakov, and Witten. On the field theory side, we derive an explicit
expression for the two point function in terms of data that may be
extracted from a DLCQ calculation at a given harmonic resolution. This
gives rise to a well defined numerical algorithm for computing the two
point function, which we test in the context of free fermions and the
't Hooft model. For the supersymmetric Yang-Mills theory with 16
supercharges that arises in the Maldacena conjecture, the algorithm is
perfectly well defined, although the size of the numerical computation
grows too fast to admit any detailed analysis at present, and our
results are only preliminary. We are, however, able to present more
detailed results on the supersymmetric DLCQ computation of the stress
energy tensor correlators for two dimensional Yang Mills theories with
(1,1) and (2,2) supersymmetries.
\end{abstract}

\vfil\vfil\vfil
\begin{flushleft}
June 1999
\end{flushleft}

\end{titlepage}
\renewcommand{\baselinestretch}{1.05}  

\section{Introduction}

There has been a great deal of excitement during this past year
following the realization that certain field theories admit concrete
realizations as a string theory on a particular background
\cite{adscft}. By now many examples of this type of correspondence for
field theories in various dimensions with various field contents have
been reported in the literature (for a comprehensive review and list
of references, see \cite{agmoo}).  However, attempts to apply these
correspondences to study the details of these theories have only met
with limited success so far. The problem stems from the fact that our
understanding of both sides of the correspondence is limited. On the
field theory side, most of what we know comes from perturbation theory
where we assume that the coupling is weak. On the string theory side,
most of what we know comes from the supergravity approximation where
the curvature is small.  There are no known situations where both
approximations are simultaneously valid. At the present time,
comparisons between the dual gauge/string theories have been
restricted to either qualitative issues or quantities constrained by
symmetry. Any improvement in our understanding of field theories
beyond perturbation theory or string theories beyond the supergravity
approximation is therefore a welcome development.

In this note we raise the Supersymmetric Discrete Light Cone Quantization
(SDLCQ) of
field theories \cite{MY,BP85,BPP} to the challenge of providing
quantitative data which can be compared against the supergravity
approximation on the string theory side of the correspondence.  We will
work in two space-time dimensions where the SDLCQ approach provides
a natural non-perturbative solution to the theory. In general, attempts to
improve the field theory side beyond perturbation theory seem like a
promising approach in two space-time dimensions where a great deal is
already known about field theories beyond perturbation theory.

We will study the field theory/string theory correspondence motivated
by considering the near-horizon decoupling limit of a D1-brane in type
IIB string theory \cite{IMSY}. The gauge theory corresponding to this
theory is the Yang-Mills theory in two dimensions with 16
supercharges.  Its SDLCQ formulation was recently reported in
\cite{DLCQ16}. This is probably the simplest known example of a field
theory/string theory correspondence involving a field theory in two
dimensions with a concrete Lagrangian formulation.

A convenient quantity that can be computed on both sides of the
correspondence is the correlation function of gauge invariant
operators \cite{GKP,Wit}. We will focus on two point functions of
the stress-energy tensor.  This turns out to be a very convenient quantity
to compute for many reasons that we will explain along the way.  Some
aspects of this as it pertains to a consideration of black hole entropy
was recently discussed in \cite{akisunny}. There are other physical
quantities often reported in the literature. In the DLCQ literature,
the spectrum of hadrons is often reported.  This would be fine for
theories in a confining phase. However, we expect the SYM in two
dimension to flow to a non-trivial conformal fixed point in the
infra-red \cite{IMSY,DVV}.  The spectrum of states will therefore form
a continuum and will be cumbersome to handle.  On the string theory
side, entropy density \cite{BISY} and the quark anti-quark potential
\cite{BISY,RY,juanwilson} are frequently reported. The definition of
entropy density requires that we place the field theory in a
space-like box which seems incommensurate with the discretized light
cone.  Similarly, a static quark anti-quark configuration does not fit
very well inside a discretized light-cone geometry.  The correlation
function of point-like operators do not suffer from these problems. We
should mention that there exists interesting work on computing the QCD
string tension \cite{adi1,adi2} directly in the field theory. These
authors find that the QCD string tension vanishes in the
supersymmetric theories which is consistent with the power law quark
anti-quark potential found on the supergravity side.

\section{Correlation functions from supergravity}

Let us begin by reviewing the computation of the correlation function
of stress energy tensors on the string theory side using the
supergravity approximation.  The computation is essentially a
generalization of \cite{GKP,Wit}.  The main conclusion on the
supergravity side was reported recently in \cite{akisunny} but we will
elaborate further on the details. The near horizon geometry of a
D1-brane in string frame takes the form
\begin{eqnarray}
ds^2& =& \alpha' \left( {U^3 \over  \sqrt{64 \pi^3 g_{YM} ^2 N}}
dx_\parallel^2 + { \sqrt{64 \pi^3 g_{YM}^2 N} \over U^{3}} dU^2 + \sqrt{64
\pi^3 g_{YM}^2 N} U d \Omega_{8-p}^2 \right) \nonumber \\
e^\phi & = & 2 \pi  g_{YM}^2 \left( {64 \pi^3 g_{YM}^2  N \over U^6}
\right)^{{1 \over 2}} .
\end{eqnarray}
In order to compute the two point function, we need to know the action
for the diagonal fluctuations around this background to the quadratic
order. What we need is an analogue of \cite{KRvN} for this background
which unfortunately is not currently available in the
literature. Fortunately, some diagonal fluctuating degrees of freedom
can be identified by following the early work on black hole absorption
cross-sections \cite{krasnitz1,krasnitz2}. In particular, we can show
that the fluctuations parameterized according to
\begin{eqnarray}
ds^2 & = & \left(1 +  f(x^0,U) +  g(x^0,U) \right) g_{00} (dx^0)^2  +
\left(1 +5 f(x^0,U) +  g(x^0,U)\right) g_{11} (dx^1)^2  \nonumber \\
&& + \left(1 +  f(x^0,U) +  g(x^0,U)\right) g_{UU} dU^2 + \left(1 +
f(x^0,U) -    {5 \over 7} g(x^0,U)\right) g_{\Omega\Omega} d \Omega_7^2
\nonumber \\
e^\phi &=& \left(1 + 3 f(x^0,U) - g(x^0,U) \right) e^{\phi_0}
\end{eqnarray}
will satisfy the equations of motion
\begin{eqnarray}
f''(U)  + {7 \over U}  f'(U) - {64 \pi^3 g_{YM}^2  N k^2 \over U^{6}} f(U)
&=&  0 \nonumber \\
g''(U)  +  {7 \over U} g'(U)- {72 \over U^2} g(U)  - {64 \pi^3 g_{YM}^2  N
k^2 \over U^6} g(U)   &=&  0 \label{fgeq}
\end{eqnarray}
by direct substitution into the equations of motion in 10
dimensions. We have assumed without loss of generality that these
fluctuation vary only along the $x^0$ direction of the world volume
coordinates like a plane wave $e^{i k x^0}$. The fields $f(U)$ and
$g(U)$ are scalars when the D1-brane is viewed as a black hole in 9
dimensions; in fact there are the minimal and the fixed scalars in
this black hole geometry. In 10 dimensions, however, we see that they
are really part of the gravitational fluctuation. We expect therefore
that they are associated with the stress-energy tensor in the operator
field correspondence of \cite{GKP,Wit}. In the case of the
correspondence between ${\cal N} = 4$ SYM and $AdS_5 \times S_5$,
superconformal invariance allowed the identification of operators and
fields in short multiplets \cite{ferrara}. For the D1-brane, we do not
have superconformal invariance and this technique is not
applicable. In fact, we expect all fields of the theory consistent
with the symmetry of a given operator to mix.  The large distance
behavior should then be dominated by the contribution with the longest
range. The field $f(k^0,U)$ appears to be the one with the longest
range since it is the lightest field.

The  equation (\ref{fgeq}) for $f(U)$ can be solved explicitly in terms of the
Bessel's function
\begin{equation}
f(U) = U^{-3}  K_{3/2} (  \sqrt{16 \pi^3 g_{YM}^2 N}  U^{-2} k  ).
\end{equation}
By thinking of $f(U)$ in direct analogy with the minimally coupled
scalar as was done in \cite{GKP,Wit}, we can compute the flux factor
\begin{equation} {\cal F} = \lim_{U_0 \rightarrow \infty} \left. {1 \over 2
\kappa_{10}^2} \sqrt{g} g^{UU} e^{-2 (\phi - \phi_{\infty})} \partial_U
\log( f(U))  \right|_{U = U_0} = {N U_0^2 k^2\over 2 g_{YM}^2} - {N^{3/2}
k^3 \over 4 g_{YM}} + \ldots
\end{equation}
up to a numerical coefficient of order one which we have suppressed.
We see that the leading non-analytic (in $k^2$) contribution is due to
the $k^3$ term, whose Fourier transform scales according
to\footnote{It is not difficult to show that for a generic $p$-brane,
$\langle {\cal O}(x){\cal O}(0) \rangle = {{N^{{7-p} \over 5-p}}
g_{YM}^{-{2 (3-p) \over 5-p}} x^{-{19+2 p - p^2 \over 5-p}}}.$}
\begin{equation}
\langle {\cal O}(x) {\cal O} (0) \rangle = {N^{{3 \over 2}} \over
g_{YM} x^5}. \label{SG}
\end{equation}
This result passes the following important consistency test.  The SYM
in 2 dimensions with 16 supercharges have conformal fixed points in
both UV and IR with central charges of order $N^2$ and $N$,
respectively. Therefore, we expect the two point function of stress
energy tensors to scale like $N^2/x^4$ and $N/x^4$ in the deep UV and
IR, respectively. According to the analysis of \cite{IMSY}, we expect
to deviate from these conformal behavior and cross over to a regime
where supergravity calculation can be trusted. The cross over occurs
at $x = 1 / g_{YM} \sqrt{N}$ and $x = \sqrt{N} / g_{YM}$. At these
points, the $N$ scaling of (\ref{SG}) and the conformal result match
in the sense of the correspondence principle \cite{garyjoe}.

\section{Correlation functions from DLCQ}

The challenge then is to attempt to reproduce the scaling relation
(\ref{SG}), fix the numerical coefficient, and determine the detail
of the cross-over behavior using SDLCQ.  Ever since the original
proposal
\cite{Dirac}, the question of equivalence between quantizing on a
light-cone and on a space-like slice have been discussed
extensively. This question is especially critical whenever a
massless particle or a zero-mode in the quantization is present.  It
is generally believed that the massless theories can be described on
the light-cone as long as we take $m\rightarrow 0$ as a limit. The
issue of zero mode have been examined by many authors. Some recent
accounts can be found in
\cite{zm1,zm2,zm3,zm4,zm5}. Generally speaking, supersymmetry seems
to save SDLCQ from complicated zero-mode issues.  We will not
contribute much to these discussions. Instead, we will formulate the
computation of the correlation function of stress energy tensor in
naive DLCQ.  To check that these results are sensible, we will first
do the computation for the free fermions and the 't Hooft model.
Extension to SYM with 16 supercharges will be essentially
straightforward, except for one caveat. In order to actually
evaluate the correlation functions, we must resort to numerical
analysis at the last stage of the computation. For the SYM with 16
supercharges, this problem grows too big too fast to be practical on
desk top computer where the current  calculations were performed.
We can only provide an algorithm,  which, when executed on an much
more powerful computer, should reproduce (\ref{SG}).
Nonetheless, the fact that we can define a concrete algorithm seems
to be a progress in the right direction.  One potential pit-fall is
the fact that the computation may not show any sign of convergence.
If this is the case, or if it converges to a result at odds with
(\ref{SG}), we must go back and re-examine the issue of equivalence
of forms and the issue of zero modes.

The technique of DLCQ is reviewed by many authors \cite{BPP,KresIgor}
so we will be brief here.  The basic idea of light-cone quantization
is to parameterize the space using light cone coordinates $x^+$ and
$x^-$ and to quantize the theory making $x^+$ play the role of time.
In the discrete light cone approach, we require the momentum $p_- =
p^+$ along the $x^-$ direction to take on discrete values in units of
$p^+/k$ where $p^+$ is the conserved total momentum of the system and
$k$ is an integer commonly referred to as the harmonic resolution.
One can think of this discretization as a consequence of compactifying
the $x^-$ coordinate on a circle with a period $2L = {2 \pi k /
p^+}$. The advantage of discretizing the light cone is the fact that
the dimension of the Hilbert space becomes finite.  Therefore, the
Hamiltonian is a finite dimensional matrix and its dynamics can be
solved explicitly.  In SDLCQ one makes the DLCQ approximation to the
supercharges and these discrete representations satisfy the
supersymmetry algebra. Therefore SDLCQ enjoys the improved
renormalization properties of supersymmetric theories.  Of course, to
recover the continuum result, we must send $k$ to infinity and as luck
would have it, we find that SDLCQ usually converges faster than the
naive DLCQ. Of course, in the process, the size of the matrices will
grow, making the computation harder and harder.

Let us now return to the problem at hand. We would like to compute a
general expression of the form
\begin{equation}
F(x^-,x^+) = \langle {\cal O}(x^-,x^+) {\cal O} (0,0) \rangle \ .
\end{equation}
In DLCQ, where we fix the total momentum in the $x^-$ direction, it is
more natural to compute its Fourier transform
\begin{equation}
\tilde{F}(P_-,x^+) = {1 \over 2 L} \langle {\cal O}(P_-,x^+) {\cal O}(-P_-,
0) \rangle\ .
\end{equation}
This can naturally be expressed in a spectrally decomposed form
\begin{equation}
\tilde{F}(P_-,x^+)= \sum_i {1 \over 2 L} \langle 0| {\cal O}(P_-) | i
\rangle e^{-i P_+^i x^+} \langle i|  {\cal O}(-P_-,0) |0 \rangle\ .
\label{master}
\end{equation}

\subsection{Free Dirac Fermions}

Let us first consider evaluating this expression for the stress-energy
tensor in the theory of free Dirac fermions as a simple example. The
Lagrangian for this theory is
\begin{equation}
{\cal L} = i \bar{\Psi} \partial \!\!\!/\, \Psi - m \bar{\Psi} \Psi
\end{equation}
where for concreteness, we take $\gamma^0 = \sigma^2, \gamma^1 = i
\sigma^1$ and we take $\Psi = 2^{-1/4} ({\psi \atop \chi})$. In terms of
the spinor components, the Lagrangian takes the form
\begin{equation}
{\cal L} = i \psi^* \partial_+ \psi + i \chi^* \partial_- \chi - {i m \over
\sqrt{2}} (\chi^* \psi - \psi^* \chi) \ .
\end{equation}
Since we treat $x^+$ as time and since $\chi$ does not have any
derivatives with respect to $x^+$ in the Lagrangian, it can be
eliminated from the equation of motion, leaving a Lagrangian which
depends only on $\psi$:
\begin{equation}
{\cal L} = i \psi^* \partial_+ \psi + i {m^2  \over 2} \psi^* { 1 \over
\partial_- } \psi \ .
\end{equation}
We can therefore express the canonical momentum and energy  as
\begin{eqnarray}
P_- & = &  \int dx^-\,
 i \psi^* \partial_- \psi \nonumber \\
P_+ & = & \int dx^-\, -{i m^2 \over 2} \psi^* { 1 \over \partial_- } \psi \ .
\end{eqnarray}
In DLCQ, we compactify $x^-$ to have period $2L$. We can then expand
$\psi$ and $\psi^*$ in modes
\begin{eqnarray}
\psi ={ 1 \over \sqrt{2L}} \left( b(n) e^{-{i n \pi  \over L} x^-} + d(-n)
e^{ { i n \pi \over L} x^-} \right) \nonumber \\
\psi^* ={ 1 \over \sqrt{2L}} \left( b(-n) e^{{i n \pi  \over L} x^-} + d(n)
e^{ -{ i n \pi \over L} x^-} \right) \ .
\end{eqnarray}
Operators $b(n)$ and $d(n)$ with positive and negative $n$ are
interpreted as a destruction and creation operators, respectively. In
a theory with only fermions, it is customary to take anti-periodic
boundary condition in order to avoid zero-mode issues. Therefore, $n$
will take on half-integer values\footnote{In SDLCQ one must use
periodic boundary condition for all the fields to preserve the
supersymmetry.}. They satisfy the anticommutation relation
\begin{equation}
\{ b(n), b(-m) \} =  \{ d(n), d(-m) \} =  \delta_{n,m} \ .
\end{equation}
Now we are ready to evaluate (\ref{master}) in DLCQ. As a simple and
convenient choice, we take
\begin{equation}
{\cal O}(-k) = {1 \over 2}\int dx^- \, \left( i \psi^* \partial_- \psi - i
(\partial_- \psi^*) \psi \right)
 e^{- {i k \pi \over L} x^-} . \label{operator}
\end{equation}
which is the Fourier transform of the local expression for $P_-$ with
the total derivative contribution adjusted to make this operator
Hermitian. Therefore, this should be thought of as the $T^{++}$
component of the stress energy tensor. For reasons that will become
clear as we go on, this turns out to be one of the simplest things to
compute. When acted on the vacuum, this operator creates a state
\begin{equation}
T^{++}(-k) |0 \rangle =   {\pi \over L} \left( { k \over 2} - n \right)
b(-k+n) d(-n) |0 \rangle  \ . \label{state}
\end{equation}
Since the fermions in this theory are free, the plane wave states
\begin{equation}
|n \rangle = b(-k+n) d(-n) |0 \rangle
\end{equation}
constitute an eigenstate. The spectrum can easily be determined by
commuting these operators:
\begin{equation}
M^2_n | n \rangle = 2 P_- P_+ | n \rangle = {m^2 }  \left( {k \over n} + {k
\over k-n} \right) | n \rangle \label{mass}
\end{equation}
which is simply the discretized version of the spectrum of a two body
continuum. All that we have to do now is calculate eigenstates of the
actual theory we are interested in and to assemble these pieces into
(\ref{master}), but we can do a little more to make the result more
presentable. The point is that since (\ref{master}) is expressed in mixed
momentum/position space notation in Minkowski space, the answer is
inherently a complex quantity that is cumbersome to display.  For the
computation of two point function, however, we can go to position space
by Fourier transforming with respect to the $L$ variable. After Fourier
transforming, it is straight forward to Euclideanize and display the two
point function as a purely real function without loosing any information.
To see how this works, let us write (\ref{master}) in the form
\begin{equation}
\tilde{F}(P_-,x^+)=
\left|{L \over \pi} \langle 0 | T^{++}(k) |n \rangle \right|^2
{1 \over 2L} {\pi^2 \over L^2} e^{-i M^2_n \over 2 ({k \pi \over L})x^+} \ .
\end{equation}
The quantity inside the absolute value sign is designed to be
independent of $L$. Now, to recover the position space form of the
correlation function, we inverse Fourier transform with respect to $P_- = k
\pi/ L$.
\begin{equation}
F(x^-,x^+)=
\left|{L \over \pi} \langle 0 | T^{++}(k) |n \rangle \right|^2
\int {d \left({k \pi\over  L}\right) \over 2 \pi}
{1 \over 2L} {\pi^2 \over L^2} e^{-i{ M^2_n \over 2 ({k \pi \over L})}x^+ -
i {k \pi \over L} x^-} .
\end{equation}
The integral over $L$ can be done explicitly and gives
\begin{equation}
F(x^-,x^+)=
\left|{L \over \pi} \langle 0 | T^{++}(k) |n \rangle \right|^2
\left({x^+ \over x^-}\right)^2 {M_n^4 \over 8 \pi^2 k^3}
K_4\left(M_n\sqrt{2 x^+ x^-}\right)
\end{equation}
where $K_4(x)$ is the 4-th modified Bessel's function. We can now
continue to Euclidean space by taking $r^2 = 2 x^+ x^-$ to be real and
considering the quantity
\begin{equation}
\left({x^- \over x^+}\right)^2 F(x^-,x^+)=
\left|{L \over \pi} \langle 0 | T^{++}(k) |n \rangle \right|^2
{M_n^4 \over 8 \pi^2 k^3} K_4(M_n r) \label{general} \ .
\end{equation}
This is a fundamental result which we will refer to a number of times
in this paper.  It has explicit dependence on the harmonic resolution
parameter $k$, but all dependence on unphysical quantities such as
the size of the circle in the $x^-$ direction and the momentum along
that direction have been canceled. For the free fermion model,
(\ref{general}) evaluates to
\begin{equation}
\left({x^- \over x^+}\right)^2 F(x^-,x^+)
= {N \over k} \sum_n  {M_n^4 \over 32 \pi^2} {(k-2n)^2 \over k^2} K_4(M_n r)
\end{equation}
with $M_n^2$ given by (\ref{mass}). The large $k$ limit can be gotten
by replacing $n \rightarrow k x$ and ${1 \over k} \sum_n \rightarrow
\int_0^1 dx$. We recover the identical result using Feynman rules. For
$r \ll m^{-1}$, this behaves like
\begin{equation}
\left({x^- \over x^+}\right)^2 F(x^-,x^+) = {N \over k} \sum_n {3 (k-2n)^2
\over 2  \pi^2 k^2 r^4} \rightarrow {N \over 2 \pi^2 r^4} \ .
\end{equation}

\subsection{'t Hooft Model}

Let us now turn to a slightly more interesting problem of computing
the correlation function of $T^{++}$ in 't Hooft's model of two
dimensional QCD \cite{thooft} in the large $N$ limit. This theory has
two characteristic scales, one determined by the mass of the quarks
and the other by the strength of the gauge coupling $g_{YM}^2 N$.  To
a large extent, this is a solvable model.  The spectrum and the wave
function of the hadrons are encoded in a one parameter integral
equation that can be handled in many ways.  A thorough analysis of
this model including the discussion of asymptotic behavior of certain
correlation functions can be found in \cite{ccg}.  This is clearly a
very mature subject.

Applying DLCQ to the 't Hooft model is tantamount to placing 't Hooft's
integral equation for the meson spectrum on a lattice. The lattice is
in the light-cone momentum space, which is precisely what is expected
when the light-cone is compactified on a circle. The DLCQ of the
't Hooft model was analyzed in detail by \cite{Horn1,Horn2}.

\begin{figure}
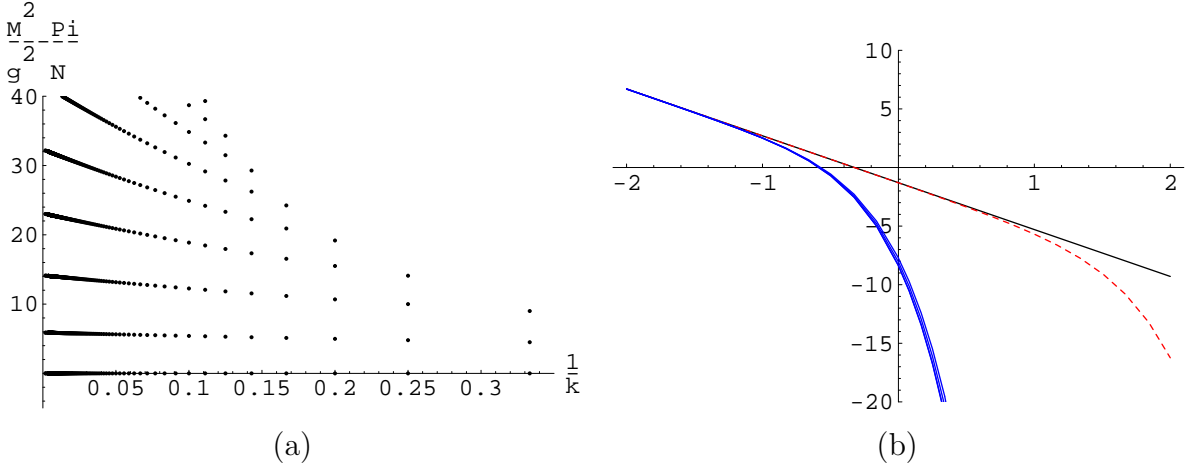

\begin{tabular}{cc}
\psfig{file=thooft.mass.epsi,width=3in}  &
\psfig{file=thooft4.epsi,width=3in}  \\
(a) & (b)
\end{tabular}
\caption{(a) Mass spectrum in units of $g_{YM}^2 N / \pi$ as a function of
$1/k$ and (b) Log-Log plot of the Correlation function ${1 \over N}
\langle T^{++}(r) T^{++}(0) \rangle$ as a function of $r$ for 't Hooft
model computed using DLCQ with $k=5$, $k=10$, $k=50$, and $k=100$ in
the units where $g_{YM} \sqrt{N / \pi} = 10^1$ and $m = 10^{-1}$. The
dotted line is for the free fermions with the same mass. \label{figa}}
\end{figure}

Our goal here is to show that the computation of (\ref{master}) is
straight forward and that it generates sensible answers. In fact,
nothing could be simpler. The 't Hooft model is nothing more than a gauged
version of the free fermion model. The Lagrangian for this theory is
simply
\begin{equation}
{\cal L} = -{1 \over 4g_{YM}^2} F^2 + i \bar{\Psi} D \!\!\!\!/\, \Psi - m
\bar{\Psi} \Psi \ .
\end{equation}
We choose the light cone gauge $A^+ = 0$ which is customary. One then
finds that the $A^-$ component of the gauge field is non-dynamical and
can be eliminated using the equation of motion, just like the $\chi$
component of the spinor in the free fermion model. Expressing
everything in terms of the only dynamical field in this theory which
is $\psi$, one finds the canonical energy and momentum operators to take
the form
\begin{eqnarray}
P_- & = & \int dx^-\,  i \psi^* \partial_- \psi \nonumber \\
P_+ & = & \int dx^-\, \left( -{i m^2 \over 2} \psi^* { 1 \over \partial_- }
\psi - {g_{YM}^2 \over 2} J_- {1 \over \partial_-^2} J_- \right)
\end{eqnarray}
where $J_- = \psi \psi^*$. All that changed in comparison to the free
fermion model is the addition of a current exchange term in the
light-cone Hamiltonian.  Therefore, all we have to do here is to
perform the identical computation specified by (\ref{general}), but
using the modified Hamiltonian, and letting $M^2_n$ and $|n\rangle$ be
the spectrum and the wavefunction of the $n$-th meson state in the
spectrum. Since in DLCQ we are always working with a finite
dimensional representation of the Hamiltonian dynamics, a small change
in the form of the Hamiltonian matrix causes no particular
difficulty. Let us discuss the result of such a computation. We will
consider the case when $g_{YM}^2 N \gg m^2$ so that the effect of the
gauge interaction is strong.  The spectrum can be computed reliably
for large $k$ and is in agreement with the results reported in
\cite{thooft} (see figure \ref{figa}.a). Since the state (\ref{state})
created by operator (\ref{operator}) is odd under parity $n
\leftrightarrow k-n$, only parity odd states contribute in the
spectral decomposition.  For $r \ll 1/g_{YM} \sqrt{N}$, we expect the
correlation function to behave just like the free fermion. The
lightest meson in the parity odd sector has a mass of order $g_{YM}
\sqrt{N}$. Due to the presence of this mass-gap, for $r \gg 1/g_{YM}
\sqrt{N}$, we expect to see an exponential damping of the correlation
function.

This is precisely the behavior we seem to be finding.  In figure
\ref{figa}.b, we illustrate the result of computing (\ref{general}) by
first constructing the mass matrix $M^2$ symbolically, then evaluating
the spectrum and the eigenfunctions numerically and assembling the
pieces.  We have chosen to set $g_{YM} \sqrt{N / \pi} = 10^1$ and $m =
10^{-1}$. We have tried harmonic resolutions $k=5$, $k=10$, $k=50$,
and $k=100$. Remarkably, the computation at a low harmonic
resolution $k$ seems not to be so far off from the result found using
larger values of $k$.  The correlation function appears to have more
or less converged by the time we reach $k=100$. We have also included
a plot for $g_{YM}^2 N = 0$ with the same mass for comparison.

One of the reasons why the convergence is relatively rapid is the fact
that the matrix element $\langle 0 | T^{++}(k) | n \rangle$ in
(\ref{general}) is not sensitive to the structure of the 't Hooft wave
function at the boundaries. To see this more clearly, recall that in
the continuum limit, we expect the eigenfunctions $|n\rangle$ to
behave as $x^\beta$ near $x=0$ where $\beta$ is determined by
\begin{equation}
0={m^2  \over g_{YM}^2 N / \pi} +  \pi \beta  \cot( \pi \beta) -1 \approx
{m^2  \over g_{YM}^2 N / \pi} - {\pi^2 \beta^2 \over 3}
\end{equation}
for small $m^2 / g_{YM}^2 N$ \cite{thooft}. When we compute matrix
elements in DLCQ approximation, we are effectively exchanging integral
expression like
\begin{equation}
\int_0^1 dx \,  a x^{a-1}  = 1
\end{equation}
by  a discretized sum
\begin{equation}
\sum_{k=1}^n    {a k^{a-1} \over n^a} \sim 1 + {a \zeta(1-a) \over n^a} +
{\cal O} ( n^{-1})
\end{equation}
whose leading correction for $a>1$ is dominated by the terms of order
$1/n$. For $a$ less than one, however, the leading correction is
controlled by terms of order $({1 \over n})^a$. In computing the form
factor for $T^{++}(k)$, we were fortunate to have only encountered an
integral whose end point behavior went as $x^{1 + \beta}$.  Had we
instead chosen to compute two point function of a scalar operators
like $\bar{\psi} \psi$ or $\bar{\psi} \gamma^5 \psi$, we would have
considered states
\begin{eqnarray}
\int dx^-\, \bar\psi \psi(x^-) e^{-{i k \pi \over L}x^-} | 0 \rangle &=&
\sum_n {m L \over 2 \pi} \left({1 \over n} - {1 \over k-n}\right) b(-n)
d(-k+n) | 0 \rangle \nonumber \\
\int dx^- \, \bar\psi \gamma^5 \psi(x^-) e^{-{i k \pi \over L}x^-}|0
\rangle &=& \sum_n {m L \over 2 \pi} \left({1 \over n} + {1 \over
k-n}\right) b(-n) d(-k+n) | 0 \rangle
\end{eqnarray}
which gives rise to a pole near $x=0$ and $x=1$ in the continuum
limit. Therefore, the matrix element in the continuum limit behaves
near $x=0$ as $x^{\beta}$.  To control the error in this case, we must
take
\begin{equation}
{\beta \over n^\beta} \ll \epsilon
\end{equation}
or equivalently
\begin{equation}
n \gg \left(\beta \over \epsilon \right) ^{1/\beta}
\end{equation}
which grows exponentially with respect to $\beta^{-1}$. We would have
had to work much harder if we had sent $m/g_{YM} \sqrt{N}$ to
zero. Luckily, this is not the case for with the $T^{++}$ operator.

\subsection{Supersymmetric Yang-Mills theory with 16 supercharges}

Finally, let us turn to the problem of computing the two point
function of the $T^{++}$ operator for the SYM with 16 supercharges.
Just as in the 't Hooft model, adopting light-cone coordinates and
choosing the light-cone gauge will eliminate the gauge boson and half
of the fermion degrees of freedom.  The most significant change comes
from the fact that the fields in this theory are in the adjoint rather
than the fundamental representations and the theory is
supersymmetric. This does not cause any fundamental problem in the
DLCQ formulation of these theories. Indeed, the SDLCQ formulation of
this \cite{DLCQ16} as well as many other related models with adjoint
fields have been studied in the literature. The main difficulty comes
from the fact that in supersymmetric theories low mass states such as
${\rm tr} [b(-n_1) b(-n_2) b(-k+n_1 + n_2)] | 0 \rangle$ with an
arbitrary number of excited quanta, or ``bits,'' appear in the
spectrum. This means that for a given harmonic resolution $k$, the
dimension of the Hilbert space grows like $\exp(\sqrt{k})$, which is roughly
the number of ways to partition $k$ into sums of integers.

The fact that the size of the problem grows  very fast is somewhat
discouraging from a numerical perspective.  Nevertheless, it is
interesting to note that DLCQ provides a well defined algorithm for
computing a physical quantity like the two point function of $T^{++}$
that can be compared with the prediction from supergravity. In the
following, we will show that this can be computed for the SYM theory
by a straight forward application of (\ref{general}), just as we saw
in the case of the 't Hooft model.

The authors of \cite{DLCQ16} have shown that the momentum operator $P^+$
is given by
\begin{equation}
P^+ = \int dx^- {\rm tr} \left[ (\partial_- X_I)^2 + i u_\alpha \partial_-
u_\alpha \right].
\end{equation}
The local Hermitian form of this operator is given by
\begin{equation}
T^{++}(x) =  {\rm tr} \left[ (\partial_- X^I)^2 + {1 \over 2} \left(i
u^\alpha \partial_- u^\alpha  - i  (\partial_- u^\alpha) u^\alpha
\right)\right], \qquad I = 1 \ldots 8, \quad \alpha = 1 \ldots 8
\end{equation}
where $X$ and $u$ are the physical adjoint scalars and fermions
respectively, following the notation of \cite{DLCQ16}.  When
discretized, these operators have the mode expansion
\begin{eqnarray}
X_{i,j}^I & = & {1 \over \sqrt{4 \pi}} \sum_{n=1}^{\infty} {1 \over
\sqrt{n}} \left[ A^I_{ij} (n) e^{-i \pi n x^-/L} + A^I_{ji}(-n) e^{i \pi n
x^-/L} \right]\nonumber \\
u_{i,j}^\alpha & = & {1 \over \sqrt{4 L}} \sum_{n=1}^{\infty} \left[
B^\alpha_{ij} (n) e^{-i \pi n x^-/L} + B^\alpha_{ji}(-n) e^{i \pi n x^-/L}
\right] .
\end{eqnarray}
In terms of these mode operators, we find
\begin{equation}
T^{++}(-k) | 0 \rangle = {\pi \over 2 L} \sum_{n=1}^{k-1}
\left[-  \sqrt{n ( k-n)}  A_{ij}(-k+n) A_{ji} (-n)  +  \left({k \over 2} -
n\right) B_{ij}(-k+n) B_{ji}(-n) \right] | 0 \rangle \label{susycorr} .
\end{equation}
Therefore, $(L/\pi) \langle 0 | T^{++}(-k) | n \rangle$ is independent
of $L$ and can be substituted directly into (\ref{general}) to give an
explicit expression for the two point function.

We see immediately that (\ref{susycorr}) has the correct small $r$
behavior, for in that limit, (\ref{susycorr}) asymptotes to (assuming
$n_b = n_f$)
\begin{equation}
\left({x^- \over x^+ }\right)^2 F(x^-,x^+) =
{N^2 \over k} \sum_n \left( {3 (k-2 n)^2 n_f \over 4 \pi^2 k^2 r^4}
+ {3 n (k-n) n_b \over  \pi^2 k^2 r^4}\right) = {N^2(2 n_b + n_f) \over 4
\pi^2 r^4}
\left(1 - {1 \over k}\right)
\end{equation}
which is what we expect for the theory of $n_b$ free bosons and $n_f$
free fermions in the large $k$ limit.

\begin{figure}
\begin{tabular}{cc}
\psfig{file=8.8.mass.epsi,width=3in}  &
\psfig{file=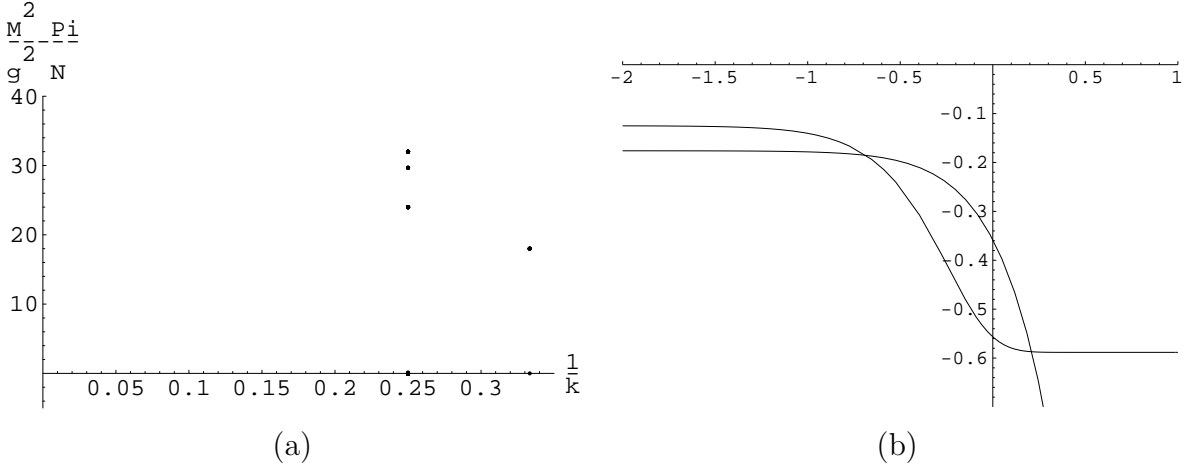,width=3in}  \\
(a) & (b)
\end{tabular}
\caption{(a) The spectrum as a function of $1/k$ and (b) the Log-Log
plot of the correlation function $\langle T^{++}(x) T^{++}(0) \rangle
\left({x^- \over x^+} \right)^2 {4 \pi^2 r^4 \over N^2 (2 n_b +n_f)}$
v.s. $r$ in the units where $g_{YM}^2 N /\pi = 1$ for $k=3$ and $k=4$.
\label{figb}}
\end{figure}

Computing this quantity beyond the small $r$ asymptotics, however,
represents a formidable technical challenge. The authors of
\cite{DLCQ16} were able to construct the mass matrix explicitly and
compute the spectrum for $k=2$, $k=3$, and $k=4$. Even for these
modest values of the harmonic resolution, the dimension of the Hilbert
space was as big as 256, 1632, and 29056 respectively (the symmetries
of the theory can be used to reduce the size of the calculation
somewhat). In figure \ref{figb}, we display results that parallel
those obtained for the 't Hooft model earlier (figure \ref{figa}) with
the currently available values of $k$, except for the fact that we
display the correlation function multiplied by a factor of $ 4 \pi^2
r^4 / N^2 (2 n_b + n_f)$, so that it now asymptotes to 1 (or 0 in the
logarithmic scale) in the $k \rightarrow \infty$ limit. In this way
any deviation from the asymptotic behavior $1/r^4$ is made more
transparent. Note that with the values of the harmonic resolution $k$
obtained at present, the spectrum in figure \ref{figb}.a is far from
resembling a dense continuum near $M=0$. Clearly, we must probe much
higher values of $k$ before we can sensibly compare our field theory
results with the prediction from supergravity.

\subsection{Supersymmetric Yang-Mills theory with less than 16 supercharges}

The computation of the correlator for the stress energy tensor in the
(8,8) model is limited by our inability to carry out the computation
for large enough harmonic resolution. It is the (8,8) model which we
are ultimately interested in solving in order to compare against the
prediction of Maldacena's conjecture in the supergravity
limit. Nevertheless, the computation of the correlation function can
just as well be applied to models with less supersymmetry.  We will
conclude by reporting the results of such a computation.

First, let us consider the theory with supercharges (1,1).  This
theory is argued not to exhibit dynamical supersymmetry breaking in
\cite{Li95,Oda97}. We can also provide a physicist's proof that
supersymmetry is not spontaneously broken for this theory by adopting
the argument of Witten for the 2+1 dimensional SYM with Chern-Simons
interaction \cite{Witten99}. In \cite{Witten99}, the index of 2+1
dimensional SYM with gauge group $SU(N)$ and 2 supercharges on $R
\times T^2$ was computed and was found to be non-vanishing for
Chern-Simons coupling $k_3 > N/2$.  If the period $L$ of one of the
circles in $T^2$ is sufficiently small, this theory is approximately
the 2-dimensional SYM with (1,1) supersymmetry with gauge coupling
$g_2^2 = g_3^2 /L$ and BF coupling $k_2 = k_3 L$ \cite{BF}. Imagine
approaching this theory by taking $L \rightarrow 0$ keeping $g_2$ and
$k_3$ fixed.  In this limit, $k_2 \rightarrow 0$ in the units of $g_2$
so the limiting theory is that of pure SYM with (1,1) supersymmetry
and a vanishing BF coupling. Choosing different values of $k_3$
corresponds to a different choice in the path of approach to this
limit. If we chose $k_3 > N/2$, we are guaranteed to have a non-zero
index for finite $L$. This means that there will be a state with zero
mass in the $L \rightarrow 0$ limit also, indicating that
supersymmetry is not spontaneously broken in this limit. On the other
hand, the index is not a well defined quantity in the $L\rightarrow 0$
limit, as a different choice of $k_3$ will lead to a different value
of the index in the $L \rightarrow 0$ limit.  In fact, the index can
be made arbitrarily large by taking $k_3$ to be also arbitrarily
large.  This suggests that there are infinitely many states forming a
continuum near $m=0$.  The index is therefore an ill defined quantity,
akin to counting the number of exactly zero energy states on a
periodic box as one takes the volume to infinity.

This theory is also believed not to be confining \cite{adi1,adi2} and
is therefore expected to exhibit non-trivial infra-red dynamics.

\begin{figure}
\begin{tabular}{cc}
\psfig{file=1.1.mass.epsi,width=3in}  &
\psfig{file=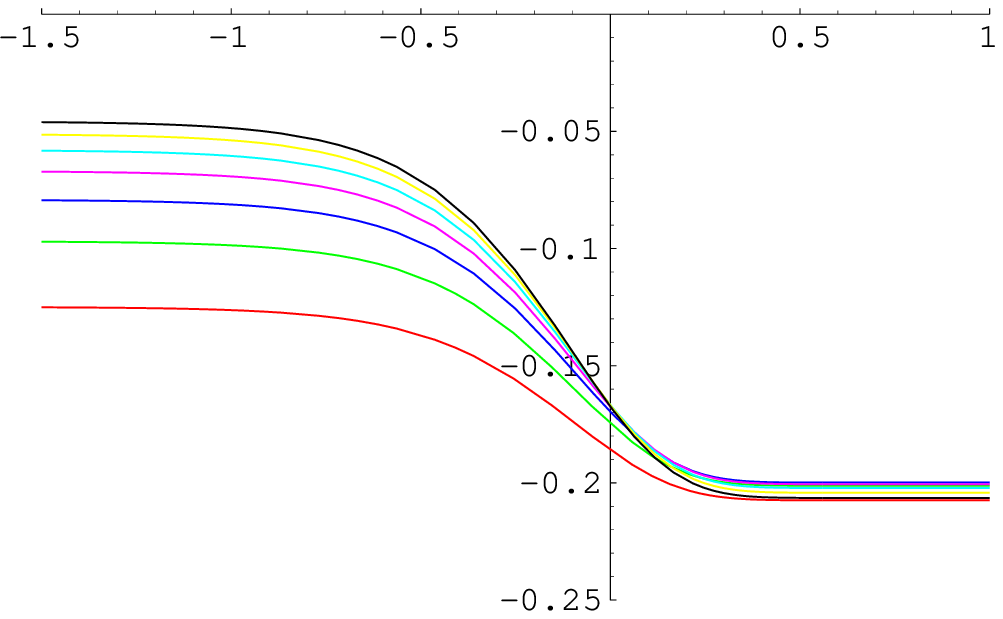,width=3in}  \\
(a) & (b)
\end{tabular}
\caption{(a) The spectrum as a function of $1/k$ and (b) the Log-Log
plot of the correlation function $\langle T^{++}(x) T^{++}(0) \rangle
\left({x^- \over x^+} \right)^2 {4 \pi^2 r^4 \over N^2 (2 n_b +n_f)}$
v.s. $r$ in the units where $g_{YM}^2 N /\pi = 1$ for $k=4 \ldots 10$.
\label{figc}}
\end{figure}

The SDLCQ of the 1+1 dimensional model with (1,1) supersymmetry
was solved in \cite{sakai,dlcq11}, and we apply these results
directly in order to compute (\ref{general}).  For simplicity, we
work to leading order in the large $N$ expansion. The spectrum of
this theory for various values of $k$, and the subsequent
computation of (\ref{general}) is illustrated in figure
\ref{figc}.a.

The spectrum of this theory at finite $k$, illustrated in figure
\ref{figc}.a, consists of $2k-2$ exactly massless states\footnote{
i.e. $k-1$ massless bosons, and their superpartners.}, accompanied
by large numbers of massive states separated by a gap. The gap appears
to be closing in the limit of large $k$ however.  We have tried
extrapolating the mass of the lightest massive state as a function of
$1/k$ by performing a least square fit to a line and a parabola, giving the
extrapolated value of $M^2 \pi / g_{YM}^2 N = 1.7$ and $M^2 \pi/g_{YM}^2 N =
-0.6$, suggesting indeed that at large $k$, the gap is closed. This is
consistent with the expectation that the spectrum is that of a
continuum starting at $M=0$ discussed earlier, although one must be
careful when the order of large $N$ and large $k$ limits are
exchanged. At finite $N$, we expect the degeneracy of $2k-2$ exactly
massless states to be broken, giving rise to precisely a continuum
of states starting at $M=0$ as expected.

\begin{figure}
\centerline{\psfig{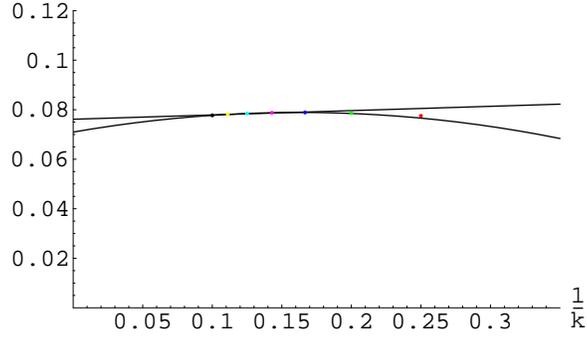}}
\caption{$ {1 \over  k^3} \sum_{n} |{L \over \pi} \langle 0 | T^{++}(k) | n
\rangle
|^2$~v.s.~$k$ from states with $M|n\rangle = 0$.  This quantity
determines the coefficient of the $1/r^4$ asymptotic tail of the
correlation function in the large $r$ limit for the (1,1) model.
\label{figd}}
\end{figure}

In the computation of the correlation function illustrated in figure
\ref{figc}.b, we find a curious feature that it asymptotes to the inverse
power law $c/r^4$ for large $r$. This behavior comes about due to the
coupling $\langle 0 | T^{++} | n \rangle$ with exactly massless states
$|n \rangle$. The contribution to (\ref{general}) from strictly massless
states are given by
\begin{equation}
\left( x^- \over x^+ \right)^2 F(x^-,x^+) = \left. \left| {L \over \pi}
\langle 0 | T^{++}(k) | n \rangle \right|^2 {M_n^4 \over 8 \pi^2 k^3 }
K_4(M_n r) \right|_{M_n =0} =
\left| {L \over \pi} \langle 0 | T^{++}(k) | n \rangle \right|^2_{M_n=0}
{6 \over k^3 \pi^2 r^4} .
\end{equation}
We have computed this quantity as a function of $1/k$ and extrapolated
to $1/k \rightarrow 0$ by fitting a line and a parabola to the
computed values for finite $1/k$. The result of this extrapolation is
illustrated in figure \ref{figd}.  The data currently available
suggests that the non-zero contribution from these massless states
persists in the large $k$ limit.

\begin{figure}
\begin{tabular}{cc}
\psfig{file=2.2.mass.epsi,width=3in}  &
\psfig{file=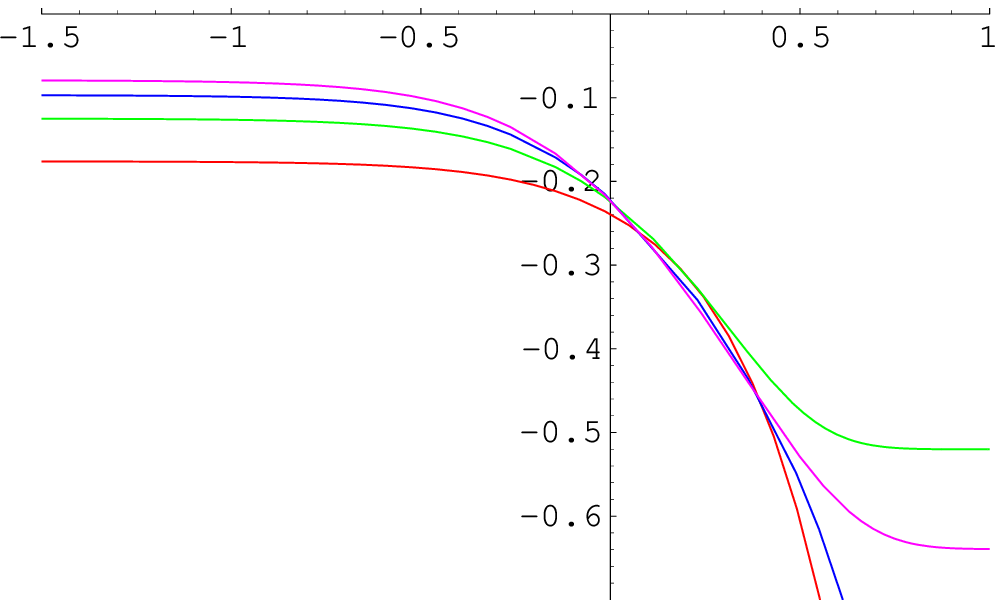,width=3in}  \\
(a) & (b)
\end{tabular}
\caption{(a) The spectrum as a function of $1/k$ and (b) the Log-Log
plot of the correlation function $\langle T^{++}(x) T^{++}(0) \rangle
\left({x^- \over x^+} \right)^2 {4 \pi^2 r^4 \over N^2 (2 n_b +n_f)}$
v.s. $r$ in the units where $g_{YM}^2 N / \pi = 1$ for $k=3 \ldots 6$.
\label{fige}}
\end{figure}

Let us now turn to the model with (2,2) supersymmetry. The SDLCQ
version of this model was solved in \cite{dlcq22}. The result of this
computation can be applied to (\ref{general}). The result is
summarized in figure \ref{fige}. This model appears to exhibit the
onset of a gapless continuum of states more rapidly than the (1,1)
model as the harmonic resolution $k$ is increased. Just as we found in
the (1,1) model, this theory contains exactly massless states in the
spectrum.  These massless states appear to couple to $T^{++} | 0
\rangle$ only for $k$ even, and the overlap appears to be decreasing
as $k$ is increased. We believe that this model is likely to exhibit a
power law behavior $c/r^\gamma$ for $\gamma > 4$ for the $T^{++}$
correlator for $r \gg g_{YM} \sqrt{N}$ in the large $N$
limit. Unfortunately, the existing numerical data do not permit the
reliable computation of the exponent $\gamma$.

\section{Conclusion}

In this article, we have provided a prescription for computing the
correlation function of the stress energy tensor $T^{++}$ in the SDLCQ
formalism, which may be readily compared with predictions provided by
a supergravity analysis following the conjecture of Maldacena. Such a
comparison requires non-perturbative methods on the field theory side,
and the SDLCQ approach appears at first sight to be particularly well
suited to this task. Unfortunately, at the present time, high enough
resolution calculation have not been made to evaluate expression
(\ref{general}) accurately enough in the case of SYM with 16
supercharges to reproduce (\ref{SG}). Significant progress is expected
however when these calculation are moved from desk top computers to a
supercomputer.  Ultimately the main obstacle will be that the number
of allowed states in the SDLCQ wavefunctions grows exponentially with
the resolution.  Nevertheless, a concrete well defined algorithm is a
great starting point for further investigations, and additional
insight may be gained by studying models with less supersymmetry, as
we have done here.

Thus, the answer to the question posed in the title is a qualified
``yes.''  Equation (\ref{general}) may be compared against (\ref{SG})
using SDLCQ, and unless other computational approaches such as the
lattice technique catches up, the SDLCQ approach remains the most
viable option.

\section*{Acknowledgments}

We thank
D.~Gross,
S.~Hellerman,
S.~Hirano,
N.~Itzhaki,
N.~Nekrassov,
Y.~Oz,
J.~Polchinski, and
E.~Witten
for illuminating discussions.  The work of AH was supported in part by
the National Science Foundation under Grant No. PHY94-07194.  The work
of FA, OL, and SP was support in part by a grant from the United
States Department of Energy.

\begingroup\raggedright\endgroup

\end{document}